\documentclass[superscriptaddress,amsmath,amssymb,aps,pra, twocolumn]{revtex4-2}

\usepackage{graphicx}% Include figure files
\usepackage{dcolumn}% Align table columns on decimal point
\usepackage{bm}% bold math
\usepackage{times}
\usepackage{hyperref}
\DeclareUnicodeCharacter{2009}{\,}

\hypersetup{
    colorlinks=true,
    linkcolor=blue,
    citecolor=blue,
    urlcolor=blue,
    pdftitle={Overleaf Example},
    pdfpagemode=FullScreen,
    }

\newcommand{\D}{\mathrm{d}}

\newcommand{\meff}{m_\mathrm{eff}}
\newcommand{\non}{\nonumber}
\newcommand{\mrm}[1]{{\mathrm{#1}}}
\newcommand{\vt}{\vphantom{\frac{1}{\sqrt{2}}}}
\usepackage{ulem}

\begin{document}

\title{How the result of a measurement of a photon's mass can turn out to be 100}

\author{Yakov Bloch}
\affiliation{Department of Physics, Bar-Ilan University, Ramat Gan 5290002, Israel}

\author{Joshua Foo}
\affiliation{Centre for Quantum Computation \& Communication Technology, School of Mathematics \& Physics, The University of Queensland, St.~Lucia, Queensland, 4072, Australia}
\affiliation{Department of Physics, Stevens Institute of Technology, Castle Point Terrace, Hoboken, New Jersey 07030, U.S.A.}

\begin{abstract}
Bohmian mechanics has garnered significant attention as an interpretation of quantum theory since the paradigmatic experiments by Kocsis et.\ al.\ [Science \textbf{332}, 6034 (2011)] and Mahler et.\ al. [Sci.\ Adv.\ \textbf{2}, 2 (2016)], which inferred the average trajectories of photons in the nonrelativistic regime. These experiments were largely motivated by Wiseman's formulation of Bohmian mechanics, which grounded these trajectories in weak measurements. Recently, Wiseman's framework was extended to the relativistic regime by expressing the velocity field of single photons in terms of weak values of the photon energy and momentum. Here, we propose an operational, weak value-based definition for the Bohmian ``local mass'' of relativistic single particles. For relativistic wavefunctions satisfying the scalar Klein-Gordon equation, this mass coincides with the effective mass defined by de Broglie in his relativistic pilot-wave theory, a quantity closely connected with the quantum potential that is responsible for Bohmian trajectory self-bending and the anomalous photoelectric effect. We demonstrate the relationship between the photon trajectories and the mass in an interferometric setup.
\end{abstract}

\maketitle

\section{Introduction}
Quantum mechanics is a useful framework for the prediction of measurement outcomes for systems delicate enough to be affected by the act of measurement \cite{albert1994quantum}. While accurately predicting the probabilities of the possible measurement outcomes and their change over time \cite{vervoort2012instrumentalist}, the theory is bound to uncertainty. Furthermore, it has been proven that a noncontextual theory that predicts observations with certainty cannot exist \cite{kochen1990problem, bell1966problem}. Although out of the ordinary theory's scope, the description of the individual quantum event has been the subject of intense research, the different aspects of which--ontological and epistemological--have been captured and scrutinized in the de Broglie-Bohm pilot wave theory, and the two-state vector formalism respectively. While the Bohmian theory suggested a hidden deterministic dynamics for the point-like corpuscle, accommodating a quantum description with the concept of particle trajectories therewith, the two-state vector formalism has led to the advent of weak measurements, probing quantum systems without causing wavefunction collapse. Merging the two, Wiseman showed \cite{Wiseman_2007} how one can in-principle obtain the deterministic Bohmian velocity field and the trajectories of nonrelativistic particles using the technique of weak measurements. The utility of his framework was demonstrated in a few notable experiments by Kocsis et.\ al.\ and Mahler et.\ al. \cite{kocsis2011observing,mahler2016experimental}, where the average trajectories of nonrelativistic particles were constructed using the weak values of the transverse component of the photon wave vector via weak measurements, which was later given an interpretation as a classical-optics effect in \cite{bliokh2013photon}. More recently, Foo et.\ al.\  \cite{Foo2022} reformulated Wiseman's nonrelativistic weak value prescription to describe the relativistic Bohmian trajectories of photons using weak values. The authors define the velocity of a Bohmian relativistic particle as the weak value of momentum, divided by the weak value of the Hamiltonian, yielding a Lorentz-covariant velocity field expressed in terms of the components of the conserved Klein-Gordon current vector. Grounding relativistic Bohmian mechanics in a measurement framework has opened the way for experimental verification of the resulting deterministic trajectories in this regime. 

Here, we build on the weak value framework of Foo et.\ al.\ by operationally defining the ``local mass'' of relativistic single photons in terms of weak values of the photon's energy and momentum. Importantly, we demonstrate that this mass is related to the one conjectured by de Broglie in his early attempts at a relativistic pilot wave theory. A measurement of the local mass would therefore constitute an experimental observation of de Broglie's effective mass. We demonstrate how Bohmian trajectory bending in a single photon interferometric setup can be understood in terms of the variable local mass that the particle acquires. 

The rest of the paper is organized as follows: in Sec.\ \ref{sec:II} and \ref{sec:III} we review the two-state vector formalism and relativistic Bohmian mechanics respectively. In Sec.\ \ref{sec:IV} we construct our weak value-based operational definition for the local photon mass, and apply this to trajectories within a Michelson-Sagnac type interferometer, simulating the trajectories in various configurations. We conclude with some final remarks in Sec.\ \ref{sec:conclusion}. 

\section{Two-state vector formalism}\label{sec:II}
A well-understood and experimentally verified feature of quantum mechanics is that a given system, prepared in a certain initial state, might end up in several different final states after measurement \cite{heisenberg1949physical}. At best, by repeatedly preparing and measuring the system, one can infer the probabilities of the possible outcomes. Therefore, ordinary quantum theory restricts itself to the prediction of these probabilities and the expression of their change over time. It has been noted \cite{aharonov1964time}, however, that the uncertainty inherent in the outcomes of quantum measurements is inconsistent with the principle of time symmetry. While the initial state is definite and predetermined, the final state is uncertain and probabilistic. Following this observation, Reznik and Aharanov proposed a time symmetrized formulation of quantum mechanics \cite{reznik1995time}. In the theory, conditions are arranged such that time symmetry with respect to preparation and measurement of a quantum system is established. Accomplished in ensembles defined by postselection as well as preselection (where both the initial and final states of the system are predetermined), time symmetrization entails fascinating opportunities for the research of extremely delicate quantum phenomena often publicized as ``paradoxes" \cite{aharonov2023time, aharonov2013quantum, aharonov2016quantum, aharonov2017case, aharonov2018extraordinary} or ``subtle interference effects" \cite{reznik2023photons, barnett2013superweak}. The theoretical framework also proves useful in the analysis of counterfactual communication protocols \cite{dressel2023counterportation}, among other applications. A remarkable property of time symmetrized systems is the weak value \cite{aharonov1990properties}, which is a measurable quantity uniquely assigned to such systems as they evolve from their initial to their final state. Expressed as, 
\begin{equation}
   \langle \hat{A}_w \rangle  = \frac{ \langle\phi|\hat{A}|\psi\rangle}{\langle\phi|\psi\rangle},
    \label{eq:Weak}
\end{equation}
where $| \psi \rangle$ is the initial and $\langle \phi|$ the final states of the system and $\hat{A}$ the quantum observable, weak values characterize a quantum system without collapsing its wavefunction. In particular, weak values generalize the concept of an expectation value to pre- and postselected ensembles, and are not restricted to the operator's spectrum--indeed, they may in general be complex. Weak values arise from a physical measurement procedure known as a weak measurement \cite{aharonov1988result},  particularly useful in sensing applications \cite{jordan2019gravitational, martinez2017ultrasensitive}, whereby the weak coupling of the pointer to the system (which is time symmetrized with respect to preparation and measurement) does not significantly perturb the system. The final state of the pointer is the result of a weak measurement, corresponding to the real part of a weak value, which collects information about the measured system without destroying coherence. Interestingly, the two-state vector formalism has been recently applied to advocate an ontological model for the existence of negative-mass particles \cite{waegell2022quantum}, facilitating positive-negative pair ``counterparticles" at the vertices of a connected graph. The latter has, in part, inspired the theory we present in the current paper, in which we delve into the epistemology of negative masses born out of the two-state vector formalism. That said, the mass we analyze in this paper has its origin elsewhere (although we do not reject the possibility of a connection between the two notions), namely that of relativistic Bohmian mechanics, which we present in what follows.

\section{Relativistic Bohmian mechanics}\label{sec:III}
Another theory accounting for individual quantum events is Bohmian mechanics \cite{bohm2006undivided,bohmPhysRev.85.166, bohm1952II, Holland}. The theory uses the wavefunction to construct a velocity field for the quantum particle in a way that is consistent with the predictions of standard quantum theory. This construction allows quantum mechanics to accommodate the notion of trajectory, implying that all dynamics can be described deterministically. To construct the velocity field, the particle’s ``Bohmian momentum,'' which is the time derivative of its hypothesized deterministic position multiplied by mass, is simply assumed to be proportional to the local wavenumber \cite{Berrylocalwave} (in this paper, for the purpose of presentation, we take the approach of "de Broglie's dynamics" as presented in \cite{colin2014instability}. The difference between ``de Broglie's dynamics" and ``Bohm's dynamics" is outside of this paper's scope since for the present purposes they are essentially the same). This notion is inherited from the theory of wave optics, and defined as the gradient of the argument of the wavefunction,
\begin{equation}
    m\frac{\D \vec{x}}{\D t} = \hbar \mathrm{Im} \{\nabla \mathrm{ln}\psi\},
    \label{eq:localwavenumber}
\end{equation} 
the dynamics of which is derived from the quantum wave equation by means of a polar decomposition of the wavefunction, 
\begin{equation}
    \psi = Re^{i S/\hbar }. 
    \label{eq:polar}
\end{equation}
The ansatz Eq.\ (\ref{eq:polar}) transforms the single complex-valued equation into a coupled pair of real ones. Specifically, in the nonrelativistic case, the Schr\"odinger equation, 
\begin{equation}
    i\hbar\  \frac{\partial\psi}{\partial t}=\ -\frac{\hbar^2}{2m}\ \mathrm{\nabla}^2 \mathrm{\psi}+V\mathrm{\psi},
    \label{eq:Schrodinger}
\end{equation}
where $m$ is the mass, and $V$ the potential term to-be-distinguished from the Bohmian ``quantum potential'' (see below), is recast into a continuity equation,
\begin{equation}
   \frac{\partial(R^2)}{\partial t}+\nabla\left(R^2\frac{\nabla S}{m}\right)=0,
    \label{eq:cont}
\end{equation}
and a Hamilton-Jacobi-like equation,
\begin{equation}
   \frac{\mathrm{\partial}S}{\mathrm{\partial }t}+\frac{\left(\nabla S\right)^2}{2m}+V-\frac{\hbar^2}{2m}\frac{\nabla^2R}{R}=0.
    \label{eq:HJ}
\end{equation}
According to the identification in (\ref{eq:localwavenumber}), the two equations specify the dynamics of the local momentum,
\begin{equation}
   \vec{p} = \nabla S.
    \label{eq:localmomentum}
\end{equation}
The second equation is called the quantum Hamilton-Jacobi equation. It differs from its classical counterpart by an extra term called the quantum potential \cite{Quantpot},
\begin{equation}
   Q=\ -\frac{\hbar^2}{2m}\frac{\nabla^2R}{R}.
    \label{eq:quantpot}
\end{equation}
In Bohmian mechanics, this term is interpreted as an
extra potential that governs the dynamics of particles and marks the departure from classical theory. The resulting quantum force (gradient of the quantum potential) is seen, in Bohmian theory, as mediating the influence of the wave on the particle, guiding it along its otherwise classical trajectory. In the Bohmian interpretation, the quantum potential accounts for all quantum phenomena. Moreover, the strength of the quantum potential does not decay with distance, a property that accounts for the nonlocality of quantum phenomena in the Bohmian perspective, and is also a necessary feature of the theory a la Bell's theorem \cite{EPRPhysRev.47.777,bellPhysicsPhysiqueFizika.1.195}. 

A suitable relativistic generalization of the local wave-vector on the right hand side of (\ref{eq:localwavenumber}) is a simple replacement of the gradient with a covariant derivative, yielding a local four-frequency,
\begin{equation}
    k^{\mu} \equiv \mathrm{Im} \{\partial^\mu \mathrm{ln}\psi\}.
    \label{eq:localfourfreq}
\end{equation}
In the context of quantum mechanics, this four-vector is proportional to the relativistic four-momentum of the particle. Since the relativistic invariant mass is the norm of the energy-momentum four vector, it is natural to define the ``local mass" \cite{superosc} to be the norm of the local energy-momentum vector,
\begin{equation}
    m^2_\mathrm{local} \equiv \frac{\hbar^2}{c^2} k_{\mu}k^{\mu}.
    \label{eq:localmass}
\end{equation}
Such a mass, proportional to the norm of the local four-frequency, is Lorentz invariant by construction.

In the Bohmian formulation of the massless scalar Klein-Gordon equation, 
\begin{equation}
    \Box \psi = 0,
    \label{eq:masslessKG}
\end{equation}
the polar decomposition of the  wavefunction (\ref{eq:polar}) uncovers the dynamics of the local four momentum. The procedure leads to a relativistic conservation equation,
\begin{equation}
   \partial_\mu (R^2 \partial^{\mu} S) = 0,
    \label{eq:continuity}
\end{equation}
where the conserved four current is given by,
\begin{equation}
   j^{\mu} = R^2 \partial^\mu S,
    \label{eq:fourcurrent}
\end{equation}
along with the relativistic quantum Hamilton-Jacobi equation,
\begin{equation}
   ( \partial_\mu S )(\partial^\mu S) = - \hbar^2  \frac{\Box R}{R},
    \label{eq:Einstein}
\end{equation}
the relativistic Hamilton-Jacobi equation for a massless particle with an extra term which is a simple relativistic generalization of the quantum potential.

Historically, a similar treatment of the scalar Klein-Gordon equation led de Broglie to define an effective mass \cite{de1987interpretation} for the relativistic Bohmian particle, in which its (classical) mass squared and its quantum potential enter on an equal footing. For a massless particle, the effective mass squared gives rise to a relativistic generalization of the quantum potential,   
\begin{equation}
   m_\mathrm{eff}^2 \equiv - \frac{\hbar^2}{c^2} \frac{\Box R}{R}.
    \label{eq:meff}
\end{equation}
With this definition, the quantum Hamilton-Jacobi equation stemming from the scalar Klein-Gordon equation (\ref{eq:Einstein}) can be elegantly rewritten as,
\begin{equation}
    m_\mathrm{local}^2 = m_\mathrm{eff}^2.
    \label{eq:meffmlocal}
\end{equation}
De Broglie used the effective mass to explain some objectionable properties of the Klein-Gordon equation in the Bohmian perspective. First, the conserved four-current (\ref{eq:fourcurrent}) might be negative in some regions of spacetime, even for positive frequency solutions \cite{Nikolic__2006,BOHM1987321,horton2002broglie,Ghose_2001,Colin_2020}. Writing the relativistic guidance equation as the simplest relativistic generalization of its classical counterpart \cite{durr2014can}, we have,
\begin{equation}
   \frac{\D x^\mu}{\D s} \propto j^{\mu},
    \label{eq:relativisticguidance}
\end{equation}
where $s$ is an affine parameter, $x^\mu$ the position four vector and $j^{\mu}$ the four current defined in (\ref{eq:fourcurrent}). When the time component of the current  is negative, the particle moves backwards in time. As a consequence, such particles might assume several positions simultaneously and move faster than $c$. Now, with the definition in (\ref{eq:meff}), writing the local mass explicitly in terms of the local momentum and energy, we arrive at the modified Einstein relation,
\begin{equation}
   E^2 = p^2c^2 + m_\mathrm{eff}^2c^4.
    \label{eq:Einsteineff}
\end{equation}
Since the effective mass squared term can be negative, not only can the photon acquire a mass, but it might also be imaginary. This is compatible with the anticipated tachyonic behaviour of relativistic particles in Bohmian mechanics. While this peculiarity is not in strict contradiction with standard quantum mechanical observations--since the latter only deals with expectation values incapable of revealing the superluminal behaviour \cite{Nikolic__2006}--the tachyonic Bohmian trajectories still pose a conceptual difficulty if one wishes to interpret them as physically real. We believe the present work provides a viable interpretation of such difficulties, as we shall explain in the next sections. Intriguingly, the relationship between the effective mass and the quantum potential has been proposed as an explanation for the anomalous photoelectric effect \cite{AnomalousPhoto}, whereby anomalous photoelectric emission and gas photo-ionization by light has been observed for single photons whose energy is lower than the work function of the material. This effect provides an experimental hint at the existence of the local mass and its effects. Just as the possibility of a negative mass particle ontology has inspired the theoretical investigations of this paper, the anomalous photoelectric effect motivates us to define a framework for the direct measurement of the local photon mass, and to compute its value for single photons. Both are accomplished in the next section.

\section{An operational definition for local mass}\label{sec:IV}

To give the local mass an operational definition, we use the language of weak values. As suggested in \cite{Foo2022}, the Bohmian velocity field of the photons can be constructed in a Lorentz covariant manner by the identification of the local momentum (in the $x$-direction), $p_x$, and energy, $E$, with weak values of the associated quantum operators, $\hat{p}_x$ and $\hat{H}$: 
\begin{align}\label{eqwvvelocity}
    v(t,x) &= c^2 \frac{\langle (\hat{p}_x)_w \rangle}{\langle \hat{H}_w \rangle},
\end{align}
where we have specialized, without loss of generality, to one spatial dimension. We note that in \cite{Berry_2012superluminal}, a similar velocity field is proposed in terms of a weak value of the operator $\hat{v} = \hat{k}/(\hat{k}+1)^{1/2}$, where $E(k) = \sqrt{k^2 + 1}$ is the relativistic dispersion relation (with $m = 1$). However a connection with a Bohmian interpretation of deterministic particle trajectories is not made.

We assume that the evolution of the state vector $| \psi (t) \rangle$ is described by the Hamiltonian $\hat{H}$ (we choose a specific form of the former when plotting trajectories in Sec.\ \ref{sec:gauss}), while postselection of the photons occurs at the position $x$ in the position eigenstate, 
\begin{align}\label{eqpositioneigenstate}
    | x \rangle &= \int\D k \: e^{-ikx} | k \rangle .  
\end{align}
From a field-theoretic point of view, Eq.\ (\ref{eqpositioneigenstate}) should be understood as the limit of a highly resolved (in space) projection onto position in a particular reference frame. Inspired by the weak value prescription given in Eq.\ (\ref{eqwvvelocity}), we use the connection between local quantum properties and weak values, to write the local mass in (\ref{eq:localmass}), as,
\begin{equation}
    m^2_\mathrm{local} c^4 \equiv ( \mathrm{Re} \langle \hat{H}_w \rangle )^2 - c^2 ( \mathrm{Re} \langle \hat{p}_w \rangle )^2 .
    \label{eq:massdef}
\end{equation}
This statement is an operational definition for the local mass of a relativistic quantum particle. We stress that this expression does not depend on the dynamical equation satisfied by the wavefunction, nor does it contain a specific form of the operators that are to be weakly measured. In what follows we implement the definition for scalar Klein-Gordon waves, understanding that this is a simplification of a full vector-valued electrodynamical theory describing photons. Nevertheless it is pertinent to study the scalar theory since the local mass coincides with de Broglie's effective mass. As suggested in \cite{Foo2022}, we can utilise the differential forms generated from the Klein-Gordon Hamiltonian and momentum to explicitly evaluate the weak values in Eq.\ (\ref{eqwvvelocity}):
\begin{align}
     {\langle \hat{H}_w \rangle} &= \frac{\langle x |\hat{H}|\psi(t)\rangle}{\langle x|\psi(t)\rangle} =  \frac{ (-i \hbar)\langle x |(\partial/\partial t)|\psi(t)\rangle}{\langle x|\psi(t)\rangle},
    \label{eq:weakenergy} 
\\
     {\langle (\hat{p}_x)_w \rangle} &= \frac{\langle x |\hat{p}|\psi(t)\rangle}{\langle x|\psi(t)\rangle} = \frac{i\hbar \langle x |(\partial/\partial x)|\psi(t)\rangle}{\langle x|\psi(t)\rangle}.
    \label{eq:weakmom}
\end{align}
In analogy to the weak value velocity field, our weak value mass can be understood as a ``classical'' expectation constructed from repeated weak measurements performed on an ensemble of identically prepared particles. 
According to (\ref{eq:meffmlocal}), when the particles are described by the scalar Klein-Gordon equation, the local mass (which is a general property of relativistic waves) is equal to the effective mass (which is unique to de Broglie's Bohmian treatment of Klein-Gordon waves). When the particle is massless, the effective mass squared is proportional to the relativistic quantum potential, the expression of which is given in (\ref{eq:meff}). In what follows we restrict ourselves to the latter case. Using the identifications in Eq.\ (\ref{eq:weakenergy}) and (\ref{eq:weakmom}), we can express the effective mass in terms of the Klein-Gordon conserved current, $j(t,x)$ and conserved current density $\rho(t,x)$:
\begin{align}
    m_\mathrm{eff}^2 c^4 &= \frac{1}{\left| \psi(t,x) \right|^4} ( \rho^2(t,x) - c^2 j^2 (t,x) ) ,
\end{align}
where $\rho(t,x) \equiv j^0(t,x)$, $j(t,x) \equiv j^1(t,x)$ are the Klein-Gordon conserved current density and conserved current respectively, with $j^\mu = 2 \mathrm{Im} \psi^\star( t,x ) \psi(t,x)$. By defining the ``(squared) effective mass density''
\begin{align}\label{eq28}
    \bar{m}_\mathrm{eff}^2 &= \meff^2 c^4 \left| \psi(t,x) \right|^4 
    \vphantom{\Big)},
\end{align}
and utilising the fact that $\rho(t,x)$ and $j(t,x)$ are components of a conserved two-vector, we find that $\bar{m}_\mathrm{eff}^2$ is an invariant quantity under Lorentz boosts, as desired.

\subsection{Gaussian Wavepackets}\label{sec:gauss}

In order to plot the time and space dependence of the effective mass, we require an explicit form of the initial state. Let us consider a similar interferometric setup as studied in \cite{Foo2022}, where the initial state is a Gaussian wavepacket of momenta, 
\begin{align}
    | \psi (t) \rangle &= \int\D k \: e^{-iE(k)t}f(k) | k \rangle 
\end{align}
where $E(k) = | k |$ corresponds to the relativistic dispersion for massless particles (noting again, the standard simplification to the scalar Klein-Gordon theory), $m = 0$, while we assume that the wavepacket itself may be prepared in a superposition of left- and right-moving momenta:
\begin{align}
    f(k) &= \mathcal{N} \bigg( \sqrt{\alpha} \exp \left[ - \frac{(k-k_0)^2}{4\sigma^2} \right] 
    \nonumber 
    \\
    & \qquad + \sqrt{1- \alpha} \exp \left[ - \frac{(k+k_0)^2}{4\sigma^2} \right] \bigg) ,
\end{align}
where $k_0$ is the centre frequency of the wavepacket(s) and $\sigma$ their bandwidth, while $\mathcal{N}$ is a normalisation constant. Using these ingredients, coupled with the optics approxiation employed in \cite{Foo2022} (wherein the magnitude of the wavevector is much larger than its spread, $k_0 \gg \sigma$), it is possible to compute explicit forms for the conserved current and conserved current density:
\begin{align}
    j(t,x) &= \beta_\mrm{R}^2 - \beta_\mrm{L}^2 + 2 \beta_\mrm{R} \beta_\mrm{L} \mathcal{S}_0 
    \vt 
    , 
    \\
    \rho(t,x) &= \beta_\mrm{R}^2 + \beta_\mrm{L}^2 + 2 \beta_\mrm{R} \beta_\mrm{L} \mathcal{T}_0 ,
    \vt 
\end{align}
where we have defined
\begin{align}
    \mathcal{S}_0 &= \frac{2\sigma^2}{k_0} \sin(2k_0x) ,
    \\
    \mathcal{T}_0 &= \cos(2k_0x) - \frac{2\sigma^2x}{k_0} \sin(2k_0x),
\end{align}
and
\begin{align}
    \beta_\mrm{R}^2 &= \alpha \sqrt{\frac{2}{\pi}} \sigma \exp \Big[-2(t-x)^2 \sigma^2 \Big]
    ,
    \\
    \beta_\mrm{L}^2 &= (1 - \alpha) \sqrt{\frac{2}{\pi}} \sigma \exp \Big[-2(t+x)^2\sigma^2\Big] . 
\end{align}
The effective mass density, Eq.\ (\ref{eq28}), is given by 
\begin{align}
    \bar{m}_\mrm{eff}^2 &= 4 \beta_\mrm{R}^2 \beta_\mrm{L}^2 \Big[1 + \mathcal{T}_0^2 - \mathcal{S}_0^2 \Big] 
    \non 
    \\
    & + 4 \beta_\mrm{L}^3 \beta_\mrm{R} \Big[ \cos(2k_0x) + \frac{2\sigma^2(t-x)}{k_0} \sin(2k_0x) \Big] 
    \non \\
    & + 4 \beta_\mrm{R}^3 \beta_\mrm{L} \Big[ \cos(2k_0x) - \frac{2\sigma^2 (t  + x) }{k_0} \sin(2k_0x) \Big].
\end{align}
In the limits $\alpha \to 0, 1$ (i.e.\ a solely right- or left-moving Gaussian), the effective mass density vanishes. Likewise for $0 < \alpha < 1$, outside the interference region, $(t \pm x) \sigma \gg 1$, the effective mass is also negligible; the photons are lightlike in this regime. Notably, the effective mass density can become negative (imaginary, upon taking the square root) in certain regions, which corresponds exactly to points at which the tangent to the velocity curve becomes spacelike (the photons becomes tachyonic). This accords with the interpretation of $\bar{m}_\mathrm{eff}^2$ as a proxy for the metric length, $m_\mathrm{eff}^2 c^4 = E^2 - p^2 c^2$, whose sign corresponds with spacelike, lightlike, and timelike spacetime intervals:
\begin{align}\label{eq36}
    \bar{m}_\mathrm{eff}^2 & \begin{cases} < 0 & \mathrm{spacelike} ,
    \\
    = 0 & \mathrm{lightlike} , 
    \\
    > 0 & \mathrm{timelike} .
    \end{cases}
\end{align}
Indeed, the branch of solutions corresponding to imaginary effective (rest) masses ($\bar{m}_\mathrm{eff}^2<0$) has long been associated with those describing relativistic tachyonic particles \cite{bilaniuk1962,BILANIUK1969}. Crucially, our measurement-based construction of the Bohmian velocity field allows, as mentioned, for the existence of ``anomalous weak values'' through which an imaginary (negative) mass (squared mass) may be inferred from weak measurements of momentum and energy. 

\begin{figure}[h]
    \centering
    \includegraphics[width=\linewidth]{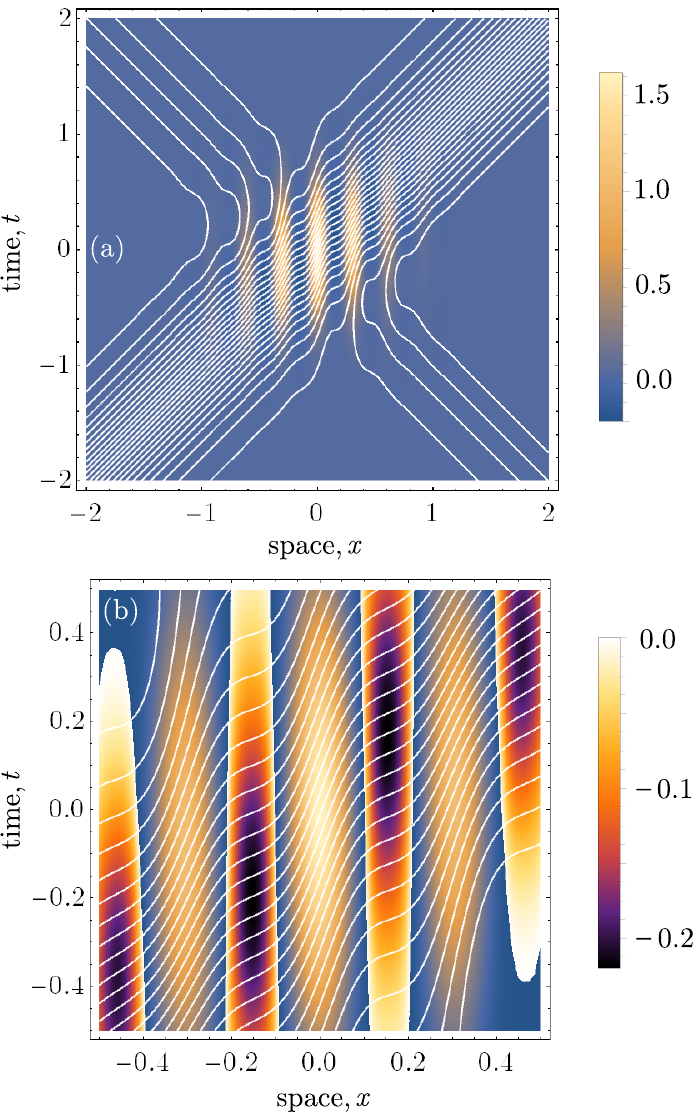}
    \caption{(a) Plot of the photon trajectories overlaid on the effective mass density, $\bar{m}_\mathrm{eff}^2$. We have used the settings $k_0/\sigma = 10$ and $\alpha = 0.83$. (b) A zoomed-in view of (a), where we have distinguished regions of positive $\bar{m}_\mathrm{eff}^2$ with negative $\bar{m}_\mathrm{eff}^2$ using a different color scheme. }
    \label{fig:1}
\end{figure}

In Fig.\ \ref{fig:1}, we have plotted the relativistic Bohmian trajectories on a spacetime diagram, with the effective mass density $\bar{m}_\mathrm{eff}^2$ underlaid. The interference fringes in the effective mass density correspond with those observed in the actual probability density, $\rho(t,x)$. Figure \ref{fig:1}(a) displays the trajectories for a Gaussian distribution of initial conditions, weighted by the Klein-Gordon density $\rho(t,x)$. As discussed, outside the interference region the effective mass density is approximately zero, corresponding to regions where the photon does not feel the influence of the quantum potential. Near the origin of coordinates (where the interference effects between the incident wavepackets become manifest) the trajectories exhibit features of single-particle interference, with the density of trajectories matching the quantum-mechanical probability density. In particular, we observe how in the regions where the photon acquires a positive effective mass, it decelerates (leading to the bunching of trajectories in these regions), while in regions of destructive interference, the tangent vector to the trajectories become spacelike, Fig.\ \ref{fig:1}(b). This behaviour is consistent with the prescription given in Eq.\ (\ref{eq36}) for the sign of the effective mass density. Likewise, when $j(t,x) = 0$ (i.e.\ the weak value of the momentum vanishes), one recovers the mass-energy equivalence for a particle in its rest frame:
\begin{align}
    E &= m_\mathrm{eff}c^2 . 
\end{align}
This accords with the association of $\rho(t,x)$ with the $\hat{T}^{00}$ component of the stress-energy tensor, commonly denoted the energy density of the (Klein-Gordon) wavefunction.

\begin{figure}[h]
    \centering
    \includegraphics[width=\linewidth]{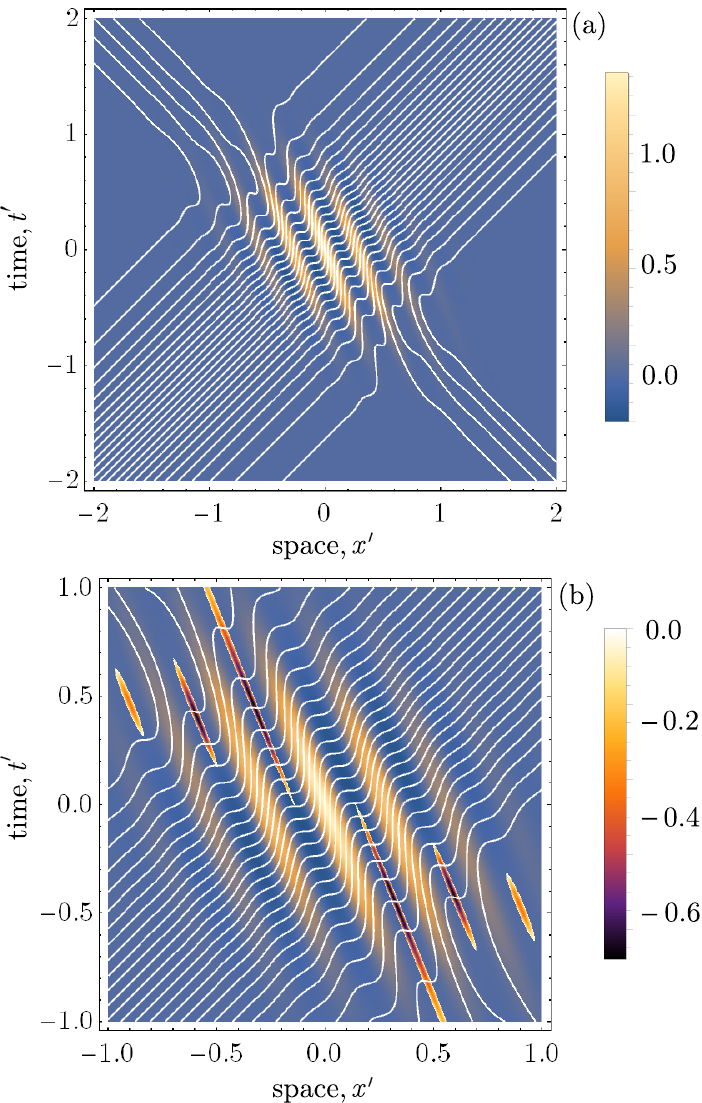}
    \caption{(a) Plot of the photon trajectories in a boosted reference frame with coordinates $(t',x')$, overlaid on the effective mass density also plotted in these coordinates. We have used the same settings as Fig.\ \ref{fig:1} with $\vartheta = 0.4$. (b) A zoomed-in view of (a), where the alternate colour scheme denotes regions where both $\bar{m}_\mathrm{eff}^2 < 0$ and $\rho(t,x) < 0$, such that trajectories travel backwards-in-time.}
    \label{fig:2}
\end{figure}

We can also consider the trajectories and corresponding effective mass under a Lorentz boost. It was shown in \cite{Foo2022} that the velocity field given by Eq.\ (\ref{eqwvvelocity}) satisfies the relativistic velocity addition rule, 
\begin{align}\label{eq37}
    v'(t,x) &= \frac{v(t,x) - \vartheta }{1 - v(t,x) \vartheta/c^2}, 
\end{align}
where $\vartheta$ is the velocity of the boosted reference frame. In Fig.\ \ref{fig:2}, we have plotted the trajectories overlaid upon the effective mass density for the same wavepacket properties as Fig.\ \ref{fig:1}, in the coordinates of the boosted observer, $(t',x') = \gamma( t - \vartheta x , x - \vartheta t)$. The Lorentz boost enacts a Doppler shift on the wavepackets, causing the width to broaden in the direction of the boost and narrow in the orthogonal direction, while the interference fringes become time-dependent due to the coupling of the spatial and temporal variables. Importantly, and as highlighed in Fig.\ \ref{fig:2}(b), the trajectories can retropropagate in the coordinates of the boosted observer (have a tangent vector pointing in the past lightcone). The segments of the trajectories that behave in this manner correspond exactly to those regions in which the above mentioned condition, $\bar{m}_\mathrm{eff}^2 < 0$ and $\rho'(t',x') < 0$, is satisfied. 

We note that such trajectories are not problematic given our interpretation of the velocity field as being a coordinate velocity obtained via measurements performed in a particular reference frame. This reference frame is originally the one in which the wavevectors and wavepacket bandwidths are equal. As shown in \cite{Foo2022}, it is always possible to find a frame in which the velocity field is globally forward-directed. On the other hand, it is inevitable that coordinate velocities with spacelike tangents will become backwards-directed upon a Lorentz boost, namely when the denominator of Eq.\ (\ref{eq37}) becomes zero. Finally, we note that unlike the early works of Refs.\ \cite{bilaniuk1962,BILANIUK1969}, and more recent studies proposing superluminal extensions to special relativity \cite{Dragan_2020,Dragan_2023}, our framework is restricted to the single-particle sector. Hence in the present work, we cannot make the standard association of retropropagating trajectories with forward-propagating antiparticle trajectories. Nevertheless, such an interpretation motivates an extension of the weak value framework to multiparticle interactions that incorporates particle production and annihilation effects.

 \section{Conclusions}\label{sec:conclusion}
In this paper, we have provided an operational definition for the local mass of a photon using the language of weak values. We connected this mass to the relativistic Bohmian trajectories of photons in an interferometric setup, and to the notion of a quantum potential arising in relativistic generalizations of Bohmian mechanics. Since standard quantum mechanics only deals with what it regards as observable quantities--namely expectation values--observing the weak, locally superluminal or locally massive photons is outside of its scope. However, using a theoretical framework which deals with individual quantum events, the aforementioned peculiar qualities of a photon can be described and even measured in principle. For the former we appeal to Bohmian mechanics, while the latter is achieved using the language of weak values. Our construction motivates novel opportunities for capturing phenomena that were, until now, out of the experimentalist's reach. Simulating such an experiment with photons, not only do we find locally massive particles of light, but they are also shown to exhibit local tachyonic behaviour, which is perceived, in a boosted frame of reference, as time-traveling particles. Surprisingly, this does not pose a challenge to relativistic principles, as the velocity we operationally define in a specific frame of reference is, in fact, a coordinate velocity which might exceed $c$. This is consistent with the fact that the locally spacelike trajectories cannot be revealed using strong measurements since no information, which the expectation values of quantum observables could reveal, travels faster than $c$. Our results are also complementary to the ``quantum metric'' interpretation given in \cite{Foo2022}, where a geometric explanation of the photon trajectories was provided. There, it was suggested that a natural generalisation of the nonrelativistic quantum potential is that of a general relativistic metric that guides the photons along geodesics. 

We hope future research utilize this approach to demonstrate more curious phenomena of this type, by way of experiment and simulation. A possible suggestion for such an experiment could be the demonstration of ``locally massless'' particles of nonzero mass. Should the experiment presented in this paper be repeated with particles of nonzero charge, an emission of Cherenkov radiation in vacuum, in regions where the particle becomes superluminal, should be expected \cite{rohrlich2002cherenkov}. Likewise, we encourage the investigation of the possibility of connection between the notion of mass explored herein and the negative mass particle ontology presented in \cite{waegell2022quantum}. Finally, since the definition used in this paper for the velocity field of relativistic quantum particles is not unique  (a notable example of an alternative definition found in \cite{Berry_2012superluminal}), the different approaches should be compared both theoretically and experimentally. One could gain an intuition of the underlying physics by plotting the different trajectories side by side. We leave this for a future work. The approach proposed herein should be applied to other relativistic equations (such as the Dirac equation) as well.
\\

\section{Acknowledgments}
We would like to thank Dr.\ Hrvoje Nikolić for useful discussions and suggestions. J.F.\ is supported by the Australian Research Council Centre of Excellence for Quantum Computation and Communication Technology (No. CE1701000012).

\bibliography{citations}

\end{document}